\documentstyle[12pt]{article}
\def\be {\begin{equation}}
\def\ee {\end{equation}}
\def\ba {\begin{eqnarray}}
\def\ea {\end{eqnarray}}

\def\bi {\begin{itemize}}
\def\ei {\end{itemize}}
\begin{document}
\def\bea{\begin{eqnarray}}
\def\eea{\end{eqnarray}}
\title{\bf {Holographic tachyon model of dark energy}}
 \author{M.R. Setare  \footnote{E-mail: rezakord@ipm.ir}
  \\ {Department of Science,  Payame Noor University. Bijar, Iran}}
\date{\small{}}
\maketitle
\begin{abstract}
In this paper we consider a correspondence between the holographic
dark energy density and tachyon energy density in FRW universe. Then
we reconstruct the potential and the dynamics of the tachyon field
which describe tachyon cosmology.
 \end{abstract}

\newpage
\section{Introduction}
Nowadays it is strongly believed that the universe is experiencing
an accelerated expansion. Recent observations from type Ia
supernovae \cite{SN} in associated with Large Scale Structure
\cite{LSS} and Cosmic Microwave Background anisotropies \cite{CMB}
have provided main evidence for this cosmic acceleration. In order
to explain why the cosmic acceleration happens, many theories have
been proposed. Although theories of trying to modify Einstein
equations constitute a big part of these attempts, the mainstream
explanation for this problem, however, is known as theories of dark
energy. It is the most accepted idea that a mysterious dominant
component, dark energy, with negative pressure, leads to this cosmic
acceleration, though its nature and cosmological origin still remain
enigmatic at present. \\
The most obvious theoretical candidate of dark energy is the
cosmological constant $\lambda$ (or vacuum energy)
\cite{Einstein:1917,cc} which has the equation of state $w=-1$. An
alternative proposal for dark energy is the dynamical dark energy
scenario.  So far, a large class of scalar-field dark energy models
have been studied, including quintessence \cite{quintessence},
K-essence \cite{kessence}, tachyon \cite{tachyon}, phantom
\cite{phantom}, ghost condensate \cite{ghost1,ghost2} and quintom
\cite{quintom}, interacting dark energy models \cite{intde},
braneworld models \cite{brane}, and Chaplygin gas models \cite{cg1},
etc.
 \\ Currently, an
interesting attempt for probing the nature of dark energy within the
framework of quantum gravity is the so-called ``holographic dark
energy'' proposal \cite{Horava:2000tb,miao}. It was shown by 'tHooft
and Susskind \cite{hologram} that effective local quantum field
theories greatly overcount degrees of freedom because the entropy
scales extensively for an effective quantum field theory in a box of
size $L$ with UV cut-off $ \Lambda$. As pointed out by \cite{myung},
attempting to solve this problem, Cohen {\it et al} showed
\cite{cohen} that in quantum field theory, short distance cut-off
$\Lambda$ is related to long distance cut-off $L$ due to the limit
set by forming a black hole. In other words the total energy of the
system with size $L$ should not exceed the mass of the same size
black hole, i.e. $L^3 \rho_{\Lambda}\leq LM_p^2$ where
$\rho_{\Lambda}$ is the quantum zero-point energy density caused by
UV cut-off $\Lambda$ and $M_P$ denotes the Planck mass (
$M_p^2=1/{8\pi G})$. The largest $L$ is required to saturate this
inequality. Then its holographic energy density is given by
$\rho_{\Lambda}= 3c^2M_p^2/8\pi L^2$ in which $c$ is a free
dimensionless parameter and coefficient 3 is for convenience. As an
application of the holographic principle in cosmology,
 it was studied by \cite{KSM} that the consequence of excluding those degrees of freedom of the system
 which will never be observed by the effective field
 theory gives rise to IR cut-off $L$ at the
 future event horizon. Thus in a universe dominated by DE, the
 future event horizon will tend to a constant of the order $H^{-1}_0$, i.e. the present
 Hubble radius.
 On the basis of the cosmological state of the holographic principle, proposed by Fischler and
Susskind \cite{fischler}, a holographic model of dark Energy (HDE)
has been proposed and studied widely in the
 literature \cite{miao,HDE}.
In the model proposed by \cite{miao}, it is discussed that
considering  the particle horizon, as the IR cut-off, the HDE
density reads
 \be
  \rho_{\Lambda}\propto a^{-2(1+\frac{1}{c})},
\ee
 that implies $w>-1/3$ which does not lead to an accelerated
universe. Also it is shown in \cite{easther} that for the case of
closed
universe, it violates the holographic bound.\\
The problem of taking apparent horizon (Hubble horizon) - the
outermost surface defined by the null rays which instantaneously are
not expanding, $R_A=1/H$ - as the IR cut-off in the flat universe
was discussed by Hsu \cite{Hsu}. According to Hsu's argument,
employing the Friedmann equation $\rho=3M^2_PH^2$ where $\rho$ is
the total energy density and taking $L=H^{-1}$ we will find
$\rho_m=3(1-c^2)M^2_PH^2$. Thus either $\rho_m$ or $\rho_{\Lambda}$
behave as $H^2$. So the DE results as pressureless, since
$\rho_{\Lambda}$ scales like matter energy density $\rho_m$ with the
scale factor $a$ as $a^{-3}$. Also, taking the apparent horizon as
the IR cut-off may result in a constant parameter of state $w$,
which is in contradiction with recent observations implying variable
$w$ \cite{varw}. On the other hand taking the event horizon, as
 the IR cut-off, gives results compatible with observations for a flat
 universe.\\
In this paper, we consider the issue of the tachyon as a source of
the dark energy. The tachyon is an unstable field which has become
important in string theory through its role in the Dirac-Born-Infeld
(DBI) action which is used to describe the D-brane action \cite
{7,8}. It has been noticed that the cosmological model based on
effective lagrangian of tachyon matter \be \label{lag}
L=-V(T)\sqrt{1-T_{,\mu}T^{,\mu}} \ee with the potential
$V(T)=\sqrt{A}$ exactly coincides with the Chaplygin gas model
\cite{{fro}, {gori}}. In the other hand, it has been pointed out
that the Chaplygin gas model can be described by a quintessence
field with well-connected potential \cite{cg1}.
 \\
  In the present paper, we suggest  a correspondence between the
holographic dark energy scenario and tachyon dark energy model. We
show this holographic description of tachyon dark energy in FRW
universe and reconstruct the potential and the dynamics of the
scalar field which describe the tachyon cosmology. The present paper
extend previous our investigations \cite{setc} in which similar
studies were done for a Chaplygin gas model.
\section{Tachyon field as holographic dark energy in flat universe}
Here we consider a four-dimensional, spatially -flat
Friedmann-Robertson-Walker universe, the Friedmann equations are as
\be \label{h1} H^2=\frac{8\pi G}{3}\rho,\ee \be
\label{h2}\frac{\ddot{a}}{a}=\frac{-4\pi G(\rho+3P)}{3} \ee where
$\rho=\rho_{NR} + \rho_{R} +\rho_{T}$ is the energy density for,
respectively, non-relativistic, relativistic and tachyon matter, and
$P$ is the corresponding pressure. We shall restrict ourselves to
consider a description of the current cosmic situation where it is
assumed that the tachyon component largely dominates and therefore
we shall disregard in what follows the non-relativistic and
relativistic components of the matter density and pressure. In this
case  the first Friedmann equation is as \be \label{h1}
H^2=\frac{8\pi G}{3}\rho_{T},\ee The energy density and pressure for
the tachyon field are as following \cite{8} \be
\label{tac}\rho_{T}=\frac{V(T)}{\sqrt{1-\dot{T}^{2}}}, \hspace {1cm}
P_{T}=-V(T)\sqrt{1-\dot{T}^{2}}\ee where $V(T)$ is the tachyon
potential energy. The barotropic index for the tachyon is \be
\label{eqes} w_{T}=\dot{T}^{2}-1 \ee Now we suggest a correspondence
between the holographic dark energy scenario and the tachyon dark
energy model.  In flat universe, our choice for holographic dark
energy density is
 \be \label{holoda}
  \rho_\Lambda=3c^2M_{p}^{2}R_{h}^{-2}.
 \ee
 where $M_{p}^{2}=\frac{1}{8\pi G}$ and
 \be \label{hor}
  R_h= a\int_t^\infty \frac{dt}{a}=a\int_a^\infty\frac{da}{Ha^2}
 \ee
 So,
 \be \label{weq} w_\Lambda=\frac{-1}{3}-\frac{2}{3c} \ee
If we establish the correspondence between the holographic dark
energy and tachyon energy density, then using
Eqs.(\ref{tac},\ref{holoda}) we have

\be \label{holoda1}
  \rho_\Lambda=3c^2M_{p}^{2}R_{h}^{-2}=\frac{V(T)}{\sqrt{1-\dot{T}^{2}}}.
 \ee
Also using Eqs.(\ref{eqes}, \ref{weq}), one can write \be
\label{weq1} w_\Lambda=\frac{-1}{3}-\frac{2}{3c}=\dot{T}^{2}-1  \ee
then \be \label{tdot} \dot{T}=\sqrt{\frac{2}{3}(1-\frac{1}{c})} \ee
We can easily obtain the evolutionary form of the tachyon field \be
\label{tac1}T=T_{0}+\sqrt{\frac{2}{3}(1-\frac{1}{c})} t \ee Now
using Eq.(\ref{holoda1}) we can obtain the tachyon potential energy
as \be \label{tacpo}V(T)=
3c^2M_{p}^{2}R_{h}^{-2}\sqrt{\frac{1}{3}(1+\frac{2}{c})}\ee If we
take $c=1$, then the behaviour of tachyon field is similar to the
cosmological constant, $\dot{T}=0$, $w_\Lambda=-1$, in this case \be
\label{tac2}T=T_{0}=constant, \hspace{1cm}
V(T)=3c^2M_{p}^{2}R_{h}^{-2} \ee The choice $c<1$ will leads to
tachyon dark energy behaving as phantom. Such a regime can be
obtained by simply Wick rotating the tachyon field so that
$T\rightarrow iT$.
\section{Tachyon field as holographic dark energy in non-flat universe}
In this section we extend the calculations of the previous section
to the non-flat universe. The first Friedmann equation is given by
\begin{equation}
\label{2eq7} H^2+\frac{k}{a^2}=\frac{1}{3M^2_p}\Big[
 \rho_{\rm \Lambda}+\rho_{\rm m}\Big].
\end{equation}
where $k$ denotes the curvature of space k=0,1,-1 for flat, closed
and open universe respectively. Define as usual
\begin{equation} \label{2eq9}\Omega_{\rm
m}=\frac{\rho_{m}}{\rho_{cr}}=\frac{ \rho_{\rm
m}}{3M_p^2H^2},\hspace{1cm} \Omega_{\rm
\Lambda}=\frac{\rho_{\Lambda}}{\rho_{cr}}=\frac{ \rho_{\rm
\Lambda}}{3M^2_pH^2},\hspace{1cm}\Omega_{k}=\frac{k}{a^2H^2}
\end{equation}
In non-flat universe, our choice for holographic dark energy density
is
 \be \label{holoda12}
  \rho_\Lambda=3c^2M_{p}^{2}L^{-2}.
 \ee
$L$ is defined as the following form\cite{nonflat}:
\begin{equation}\label{leq}
 L=ar(t),
\end{equation}
here, $a$, is scale factor and $r(t)$ is relevant to the future
event horizon of the universe. Given the fact that
\begin{eqnarray}
\int_0^{r_1}{dr\over \sqrt{1-kr^2}}&=&\frac{1}{\sqrt{|k|}}{\rm
sinn}^{-1}(\sqrt{|k|}\,r_1)\nonumber\\
&=&\left\{\begin{array}{ll}
\sin^{-1}(\sqrt{|k|}\,r_1)/\sqrt{|k|},\ \ \ \ \ \ &k=1,\\
r_1,&k=0,\\
\sinh^{-1}(\sqrt{|k|}\,r_1)/\sqrt{|k|},&k=-1,
\end{array}\right.
\end{eqnarray}
one can easily derive \be \label{leh} L=\frac{a(t) {\rm
sinn}[\sqrt{|k|}\,R_{h}(t)/a(t)]}{\sqrt{|k|}},\ee where $R_h$ is the
future event horizon given by (\ref{hor}). By considering  the
definition of holographic energy density $\rho_{\rm \Lambda}$, one
can find \cite{{set1},{set2}}:
\begin{equation}\label{stateq}
w_{\rm \Lambda}=-[\frac{1}{3}+\frac{2\sqrt{\Omega_{\rm
\Lambda}}}{3c}\frac{1}{\sqrt{|k|}}\rm cosn(\sqrt{|k|}\,R_{h}/a)].
\end{equation}
 where
\begin{equation}
\frac{1}{\sqrt{|k|}}{\rm cosn}(\sqrt{|k|}x)
=\left\{\begin{array}{ll}
\cos(x),\ \ \ \ \ \ &k=1,\\
1,&k=0,\\
\cosh(x),&k=-1.
\end{array}\right.
\end{equation}
If again we establish the correspondence between the holographic
dark energy and tachyon energy density, then using
Eqs.(\ref{tac},\ref{holoda12}) we have

\be \label{holoda123}
  \rho_\Lambda=3c^2M_{p}^{2}L^{-2}=\frac{V(T)}{\sqrt{1-\dot{T}^{2}}}.
 \ee
 Also using Eqs.(\ref{eqes}, \ref{stateq}), one can write \be
\label{weq1} w_\Lambda=-[\frac{1}{3}+\frac{2\sqrt{\Omega_{\rm
\Lambda}}}{3c}\frac{1}{\sqrt{|k|}}\rm
cosn(\sqrt{|k|}\,R_{h}/a)]=\dot{T}^{2}-1  \ee then \be \label{tdot}
\dot{T}=\sqrt{\frac{2}{3}[1-\frac{\sqrt{\Omega_{\rm
\Lambda}}}{c}\frac{1}{\sqrt{|k|}}\rm cosn(\sqrt{|k|}\,R_{h}/a)]} \ee
Now using Eq.(\ref{holoda123}) we can obtain the tachyon potential
energy as \be \label{tacpo1}V(T)=
3c^2M_{p}^{2}L^{-2}\sqrt{\frac{1}{3}[1+\frac{2\sqrt{\Omega_{\rm
\Lambda}}}{c}\frac{1}{\sqrt{|k|}}\rm cosn(\sqrt{|k|}\,R_{h}/a)]}\ee
Differenating Eq.(\ref{2eq7}) with respect to the cosmic time $t$,
one find \be \label{hdot}\dot{H}=\frac{\dot{\rho}}{6H
M_{p}^{2}}+\frac{k}{a^2} \ee where $\rho=\rho_{m}+\rho_{\Lambda}$ is
the total energy density. Now we write continuity equation for dark
energy and cold dark matter as \be \label{doro}
\dot{\rho}=-3H(1+w)\rho \ee where \be
\label{weq}w=\frac{w_{\Lambda}\rho_{\Lambda}}{\rho}=\frac{\Omega_{\Lambda}w_{\Lambda}}{1+\frac{k}{a^2H^2}}
\ee Substitute $\dot{\rho}$ into Eq.(\ref{hdot}), we obtain \be
\label{weq2}
w=\frac{2/3(\frac{k}{a^2}-\dot{H})}{H^2+\frac{k}{a^2}}-1 \ee Using
Eqs.(\ref{weq}, \ref{weq2}), one can rewrite the holographic energy
equation of state as \be \label{eqes1}
w_{\Lambda}=\frac{-1}{3\Omega_{\Lambda}H^{2}}(2\dot{H}+3H^2+\frac{k}{a^2})
\ee Therefore one can rewrite Eqs.(\ref{tdot},\ref{tacpo1})
respectively as \be \label{kintac}\dot{T}^{2}=1-
\frac{1}{3\Omega_{\Lambda}H^{2}}(2\dot{H}+3H^2+\frac{k}{a^2})\ee \be
\label{pottac}V(T)=\frac{3c^2M_{p}^{2}}{H
L^{2}}\sqrt{\frac{2\dot{H}+3H^2+\frac{k}{a^2}}{3\Omega_{\Lambda}}}
\ee Using definitions
$\Omega_{\Lambda}=\frac{\rho_{\Lambda}}{\rho_{cr}}$ and
$\rho_{cr}=3M_{p}^{2}H^2$, we get

\begin{equation}\label{hl}
HL=\frac{c}{\sqrt{\Omega_{\Lambda}}}
\end{equation}
then
 \be
\label{pottac1}V(T)=HM_{p}^{2}\sqrt{3\Omega_{\Lambda}
(2\dot{H}+3H^2+\frac{k}{a^2})} \ee In similar to the \cite {{odi1},
{odi2}, {mset}}, we can define $\dot{T}^{2}$ and $V(T)$ in terms of
single function $f(T)$ as \be \label{kintac1}-1=1-
\frac{1}{3\Omega_{\Lambda}f^{2}(T)}[2f'(T)+3f^2(T)+\frac{k}{a^2}]\ee
\be \label{pottac11}V(T)=f(T)M_{p}^{2}\sqrt{3\Omega_{\Lambda}
[2f'(T)+3f^2(T)+\frac{k}{a^2}]} \ee Hence, the following solution
are obtained \be \label{sol} T=it, \hspace{1cm} H=f(it) \ee From
Eq.(\ref{kintac1}) we get \be
\label{keq}\frac{k}{a^2}=3f^2(T)(2\Omega_{\Lambda}-1)-2f'(T)\ee
Substitute the above $\frac{k}{a^2}$ into Eq.(\ref{pottac11}), we
obtain the tachyon potential as \be
\label{pottac111}V(T)=3\sqrt{2}M_{p}^{2}\Omega_{\Lambda} f^2(T) \ee
One can check that the solution (\ref{sol}) satisfies the following
tachyon field equation \be \label{tacequation}
\frac{\ddot{T}}{1-\dot{T}^{2}}+3H\dot{T}+\frac{V'}{V}=0 \ee
Therefore by the above condition and using Eq(\ref{pottac111}),
$f(T)$ in our model must satisfy following relation \be
\label{coneq} 3if(T)+\frac{2f'(T)}{f(T)}=0\ee Elementary algebra now
gives the $f(T)$ to be of the form \be \label{fequa}f(T)=\frac{2}{3i
T} \ee In this case, we can determine the potential to be\be
\label{pottac1110}V(T)=\frac{-4\sqrt{2}}{3}M_{p}^{2}\Omega_{\Lambda}\frac{1}{T^{2}}
 \ee For the tachyon self-interaction, there are a number of models
 which one can consider, some being motivated by non-perturbative
 string theory and others purely by phenomenology. The authors of
 \cite{cg} have studied a wide renge of potentials, they have shown that in the presence of a tachyon
 field $T$ with potential $V(T)$  and a barotropic perfect fluids,
 the cosmological dynamics depends on the asymptotic behavior of the
 quantity $\lambda=\frac{-M_{p} V'}{V^{3/2}}$. If $\lambda$ is a
 constant, which corresponds to an inverse square potential $V(T)\propto
 T^{-2}$,
 there exists one stable critical point that gives an acceleration
 of the universe at late times. \footnote{$\Omega_\Lambda$ is not constant, the differential equation for $\Omega_{\Lambda}$ is
\be \label{evol}
\frac{d\Omega_{\Lambda}}{dx}=\frac{\dot{\Omega_{\Lambda}}}{H}=3\Omega_{\Lambda}(1+\Omega_{k}-\Omega_{\Lambda})
[\frac{1}{3}+\frac{2\sqrt{\Omega_{\rm
\Lambda}}}{3c}\frac{1}{\sqrt{|k|}}\rm cosn(\sqrt{|k|}\,R_{h}/a)] \ee
where $x=\ln a$. The above equation describes the behavior of the
holographic dark energy completely, in the spatially flat case, i.e,
$k=0$:\be \label{evol}
\frac{d\Omega_{\Lambda}}{dx}=\frac{\dot{\Omega_{\Lambda}}}{H}=\Omega_{\Lambda}(1+\Omega_{k}-\Omega_{\Lambda})
[1+\frac{2\sqrt{\Omega_{\rm \Lambda}}}{c}] \ee

 it can be solved exactly for arbitrary c, the solution for $c = 1$ \cite{miao} is
 as following
 \be \label{sol1}
 \ln\Omega_{\Lambda}-\frac{1}{3}(1-\sqrt{\Omega_{\Lambda}})+(1+2
 \sqrt{\Omega_{\Lambda}})-\frac{8}{3}(1+
 \sqrt{\Omega_{\Lambda}})=\ln a+x_{0}\ee where $x_{0}$ fixed by L.H.S. of (\ref{sol1})
 with $\Omega_{\Lambda}$ and $a$ replaced with present time values.
 From Eq.(\ref{sol}) one can see that $\Omega_{\Lambda}$ depend to $T$,
 therefore, the potential (\ref{pottac1110}) does not only vary
as $T^{-2}$, but at late time $\Omega_{\Lambda}$ increases to $1$.
Then it is interesting that in our model, at late time where
$\Omega_{\Lambda}=1$, we obtain $V(T)\propto
 T^{-2}$, similar to the result of \cite{cg}. In fact only in this case potential is
as $V(T)$. The additional dependence through $\Omega_{\Lambda}$
makes it dependent not only on the tachyon field but through $H^2$,
on the matter component as well.} With this result we can claim that
only the
 potentials which have the above form are consistent with the
 holographic approach of tachyon dark energy model. The property of
 the holographic dark energy is strongly depend on the parameter
 $c$.  From Eqs.(\ref{eqes1}), (\ref{pottac1}) we have
\be \label{pottacho}V(T)=HM_{p}^{2}\sqrt{-9\Omega_{\Lambda}^{2} H^2
w_{\Lambda}}=3 H^{2}M_{p}^{2}\Omega_{\Lambda}\sqrt{- w_{\Lambda}}\ee
Substitute $w_{\Lambda}$ into the above equation \be
\label{pottacho1}V(T)=3
H^{2}M_{p}^{2}\Omega_{\Lambda}\sqrt{[\frac{1}{3}+\frac{2\sqrt{\Omega_{\rm
\Lambda}}}{3c}\frac{1}{\sqrt{|k|}}\rm cosn(\sqrt{|k|}\,R_{h}/a)]}\ee
In the flat case we have\be \label{pottacho11}V(T)=3
H^{2}M_{p}^{2}\Omega_{\Lambda}\sqrt{\frac{1}{3}(1+\frac{2\sqrt{\Omega_{\rm
\Lambda}}}{c})}\ee which is exactly Eq.(15) in \cite{zhang}.
\footnote{This reference appeared on the arXiv by the number
arXiv:0706.1185 [astro-ph] after my submission to the arXxiv with
number  arXiv:0705.3517 [hep-th]. } The holographic dark energy
model has been tested and constrained by various astronomical
observations, in both flat and non-flat cases. These observational
data include type Ia supernovae, cosmic microwave background, baryon
acoustic oscillation, and the X-ray gas mass fraction of galaxy
clusters. According to the analysis of the observational data for
the holographic dark energy model, we find that generally $c<1$, and
the holographic dark energy thus behaves like a quintom-type dark
energy. When including the spatial curvature contribution, the
fitting result shows that the closed universe is marginally favored.
For the closed universe the equation of state is given by
\footnote{Flat case is discussed in \cite{zhang}} \be \label{cols}
w_{\Lambda}=\frac{-1}{3}(1+\frac{2\sqrt{\Omega_{\rm
\Lambda}}}{c}\cos x) \ee then $\frac{-1}{3}(1+\frac{2}{c})\leq
w_{\Lambda}\leq \frac{-1}{3}(1-\frac{2}{c})$, when $0\leq
\Omega_{\rm \Lambda}\leq1$. Similar to the flat case, when $c<1$,
the equation of state cross $w_{\Lambda}=-1$ (from $w_{\Lambda}>-1$
evolves to $w_{\Lambda}<-1$ ). When $c> 2$, $w_{\Lambda}$ evolve in
the region $-1<w_{\Lambda}<0$. Since the equation of state of
tachyon field also evolves in this region, then one can say, in the
closed universe case we can consider a correspondence between the
holographic dark energy density and tachyon energy density if $c>
2$. If we take $c=1$, and taking $\Omega_{\Lambda}=0.73$ for the
present time, the lower bound of $w_{\rm \Lambda}$ is $-0.9$.
Therefore it is impossible to have $w_{\rm \Lambda}$ crossing $-1$.
This implies that one can not generate phantom-like equation of
state from an holographic dark energy model with  $c=1$ in non-flat
universe. In the other hand as we have shown previously with the
choice of $c\leq 0.84$, the interacting holographic dark energy can
be described  by a phantom scalar field \cite{mset}.  Therefore the
parameter $c$ plays a crucial role in the model.

\section{Conclusions}
Based on cosmological state of holographic principle, proposed by
Fischler and Susskind \cite{fischler}, the Holographic model of Dark
Energy (HDE) has been proposed and studied widely in the
 literature \cite{miao,HDE}. In \cite{HG} using the type Ia
 supernova data, the model of HDE is constrained once
 when c is unity and another time when c is taken as free
 parameter. It is concluded that the HDE is consistent with recent observations, but future observations are needed to
 constrain this model more precisely.\\
Within the different candidates to play the role of the dark energy,
tachyon, has emerged as a possible source of dark energy for a
particular class of potentials \cite{tac}.
\\
In this paper we have associated the holographic dark energy in FRW
universe with a tachyon field which describe the tachyon cosmology.
We have shown that the holographic dark energy can be described  by
the tachyon field in a certain way. Then a correspondence between
the holographic dark energy and tachyon model of dark energy has
been established, and the potential of the holographic tachyon field
and the dynamics of the field have been reconstructed. For the
holographic tachyon model constructed in the present paper, the
tachyon potential can be determined by Eq.(\ref{pottac1110}). We saw
that the parameter $c$ plays a crucial role in the model: $c\geq 1$
makes the holographic dark energy behave as quintessence-type dark
energy with $w_{\Lambda}\geq -1$, and $c<1$ makes the holographic
dark energy behave as quintom-type dark energy with $w_{\Lambda}$
crossing $-1$ during the evolution history. Hence, we see, the
determining of the value of $c$ is a key point to the feature of the
holographic dark energy and the ultimate fate of the universe as
well. However, in the recent fit studies, different groups gave
different values to $c$. A direct fit of the present available SNe
Ia data with this holographic model indicates that the best fit
result is $c=0.21$ \cite{HG}. Recently, by calculating the average
equation of state of the dark energy and the angular scale of the
acoustic oscillation from the BOOMERANG and WMAP data on the CMB to
constrain the holographic dark energy model, the authors show that
the reasonable result is $c\sim 0.7$ \cite{cmb1}. In the other hand,
in the study of the constraints on the dark energy from the
holographic connection to the small $l$ CMB suppression, an opposite
result is derived, i.e. it implies the best fit result is $c=2.1$
\cite{cmb3}.


\begin{thebibliography}{99}
\bibitem{SN}
  A.~G.~Riess {\it et al.}  [Supernova Search Team Collaboration],
  Astron.\ J.\  {\bf 116}, 1009 (1998)
  [astro-ph/9805201];\\
  S.~Perlmutter {\it et al.}  [Supernova Cosmology Project Collaboration],
  Astrophys.\ J.\  {\bf 517}, 565 (1999)
  [astro-ph/9812133];\\
  P.~Astier {\it et al.},
  Astron.\ Astrophys.\  {\bf 447}, 31 (2006)
  [astro-ph/0510447].

\bibitem{LSS}
  K.~Abazajian {\it et al.}  [SDSS Collaboration],
  Astron.\ J.\  {\bf 128}, 502 (2004)
  [astro-ph/0403325];\\
  K.~Abazajian {\it et al.}  [SDSS Collaboration],
  Astron.\ J.\  {\bf 129}, 1755 (2005)
  [astro-ph/0410239].

\bibitem{CMB}
  D.~N.~Spergel {\it et al.}  [WMAP Collaboration],
  Astrophys.\ J.\ Suppl.\  {\bf 148}, 175 (2003)
  [astro-ph/0302209];\\
  D.~N.~Spergel {\it et al.},
  astro-ph/0603449.
  \bibitem{Einstein:1917} A. Einstein, Sitzungsber. K. Preuss.
Akad. Wiss. 142 (1917) [{\it The Principle of Relativity} (Dover,
New York, 1952), p. 177].

\bibitem{cc}S.~Weinberg,
Rev.\ Mod.\ Phys.\  {\bf 61} 1 (1989);\\
  V.~Sahni and A.~A.~Starobinsky,
  Int.\ J.\ Mod.\ Phys.\ D {\bf 9}, 373 (2000)
  [astro-ph/9904398];\\
S.~M.~Carroll,
Living\ Rev.\ Rel.\ {\bf 4} 1 (2001) [astro-ph/0004075];\\
P.~J.~E.~Peebles and B.~Ratra,
Rev.\ Mod.\ Phys.\  {\bf 75} 559 (2003) [astro-ph/0207347];\\
T.~Padmanabhan,
Phys.\ Rept.\  {\bf 380} 235 (2003) [hep-th/0212290].
\bibitem{quintessence}
  P.~J.~E.~Peebles and B.~Ratra,
  Astrophys.\ J.\  {\bf 325} L17 (1988);\\
  B.~Ratra and P.~J.~E.~Peebles,
  Phys.\ Rev.\ D {\bf 37} 3406 (1988);\\
  C.~Wetterich,
  Nucl.\ Phys.\ B {\bf 302} 668 (1988);\\
  J.~A.~Frieman, C.~T.~Hill, A.~Stebbins and I.~Waga,
  Phys.\ Rev.\ Lett.\  {\bf 75}, 2077 (1995)
  [astro-ph/9505060];\\
  M.~S.~Turner and M.~J.~White,
  Phys.\ Rev.\ D {\bf 56}, 4439 (1997)
  [astro-ph/9701138];\\
  R.~R.~Caldwell, R.~Dave and P.~J.~Steinhardt,
  Phys.\ Rev.\ Lett.\  {\bf 80}, 1582 (1998)
  [astro-ph/9708069];\\
  A.~R.~Liddle and R.~J.~Scherrer,
  Phys.\ Rev.\ D {\bf 59}, 023509 (1999)
  [astro-ph/9809272];\\
  I.~Zlatev, L.~M.~Wang and P.~J.~Steinhardt,
  Phys.\ Rev.\ Lett.\  {\bf 82}, 896 (1999)
  [astro-ph/9807002];\\
  P.~J.~Steinhardt, L.~M.~Wang and I.~Zlatev,
  Phys.\ Rev.\ D {\bf 59}, 123504 (1999)
  [astro-ph/9812313];
Z. G. Huang,  H. Q. Lu, and W. Fang, Class. Quant. Grav. {\bf23},
6215, (2006), [hep-th/0604160 ]

\bibitem{kessence}
  C.~Armendariz-Picon, V.~F.~Mukhanov and P.~J.~Steinhardt,
  Phys.\ Rev.\ Lett.\  {\bf 85}, 4438 (2000)
  [astro-ph/0004134];\\
  C.~Armendariz-Picon, V.~F.~Mukhanov and P.~J.~Steinhardt,
  Phys.\ Rev.\ D {\bf 63}, 103510 (2001)
  [astro-ph/0006373].

\bibitem{tachyon}
  A.~Sen,
  JHEP {\bf 0207}, 065 (2002)
  [hep-th/0203265].
\bibitem{phantom}
  R.~R.~Caldwell,
  Phys.\ Lett.\ B {\bf 545}, 23 (2002)
  [astro-ph/9908168];\\
  R.~R.~Caldwell, M.~Kamionkowski and N.~N.~Weinberg,
  Phys.\ Rev.\ Lett.\  {\bf 91}, 071301 (2003)
  [astro-ph/0302506]\\
 S. Nojiri and S. D.
Odintsov, Phys. Lett., B {\bf562}, 147, (2003);\\
S. Nojiri and S. D. Odintsov, Phys. Lett., B {\bf565}, 1, (2003).
\bibitem{ghost1}
  N.~Arkani-Hamed, H.~C.~Cheng, M.~A.~Luty and S.~Mukohyama,
  JHEP {\bf 0405}, 074 (2004)
  [hep-th/0312099].


\bibitem{ghost2}
  F.~Piazza and S.~Tsujikawa,
  JCAP {\bf 0407}, 004 (2004)
  [hep-th/0405054].


\bibitem{quintom}
  B.~Feng, X.~L.~Wang and X.~M.~Zhang,
  Phys.\ Lett.\ B {\bf 607}, 35 (2005)
  [astro-ph/0404224];\\
  Z.~K.~Guo, Y.~S.~Piao, X.~M.~Zhang and Y.~Z.~Zhang,
  Phys.\ Lett.\ B {\bf 608}, 177 (2005)
  [astro-ph/0410654];\\
  X.~Zhang,
  Commun.\ Theor.\ Phys.\  {\bf 44}, 762 (2005);\\
  A. Anisimov, E. Babichev and A. Vikman,
  JCAP {\bf 0506}, 006 (2005)
  [astro-ph/0504560];\\
M. R. Setare, Phys. Lett. B {\bf641}, 130, (2006);\\
E. Elizalde , S. Nojiri, and S. D. Odintsov, Phys. Rev. {\bf D70},
043539, (2004);\\
S. Nojiri, S. D. Odintsov, and S. Tsujikawa, Phys. Rev. {\bf D71},
063004, (2005).

\bibitem{intde}
  L.~Amendola,
  Phys.\ Rev.\ D {\bf 62}, 043511 (2000)
  [astro-ph/9908023];\\
  D.~Comelli, M.~Pietroni and A.~Riotto,
  Phys.\ Lett.\ B {\bf 571}, 115 (2003)
  [hep-ph/0302080];\\
  X.~Zhang,
  Mod.\ Phys.\ Lett.\ A {\bf 20}, 2575 (2005)
  [astro-ph/0503072]\\
M. Szydlowski, Phys. Lett. B {\bf632}, 1 (2006),
[astro-ph/0502034];\\
M. Szydlowski, A. Kurek, and A. Krawiec Phys. Lett. B{\bf642}, 171,
(2006) [astro-ph/0604327].
\bibitem{brane}
  C.~Deffayet, G.~R.~Dvali and G.~Gabadadze,
  Phys.\ Rev.\ D {\bf 65}, 044023 (2002)
  [astro-ph/0105068];\\
  V.~Sahni and Y.~Shtanov,
  JCAP {\bf 0311}, 014 (2003)
  [astro-ph/0202346].

\bibitem{cg1}
  A.~Y.~Kamenshchik, U.~Moschella and V.~Pasquier,
  Phys.\ Lett.\ B {\bf 511}, 265 (2001)
  [gr-qc/0103004].

\bibitem{Horava:2000tb}
  P.~Horava and D.~Minic,
  Phys.\ Rev.\ Lett.\  {\bf 85}, 1610 (2000)
  [hep-th/0001145];\\
  S.~D.~Thomas,
  Phys.\ Rev.\ Lett.\  {\bf 89}, 081301 (2002).

\bibitem{miao}
  M.~Li,
  Phys.\ Lett.\ B {\bf 603}, 1 (2004)
  [hep-th/0403127].
\bibitem{hologram}G.~'t Hooft,
  gr-qc/9310026;\\ L. Susskind, J. Math. Phys, {\bf36}, (1995),
6377-6396.
\bibitem{myung}  Y. S. Myung, Phys. Lett. B {\bf610}, (2005), 18-22.
\bibitem{cohen} A. Cohen, D. Kaplan and A. Nelson, Phys. Rev. Lett
82, (1999), 4971.
\bibitem{KSM} K. Enqvist, S. Hannestad and M. S. Sloth, JCAP, 0502, (2005)
004.
\bibitem{fischler} W. Fischler and L. Susskind, hep-th/9806039.


\bibitem{HDE} D. N. Vollic, hep-th/0306149; H. Li, Z. K.
Guo and Y. Z. Zhang, astro-ph/0602521; J. P. B. Almeida and J. G.
Pereira, gr-qc/0602103; D. Pavon and W. Zimdahl, hep-th/0511053; Y.
Gong, Phys. Rev., D, 70, (2004), 064029; B. Wang, E. Abdalla, R. K.
Su, Phys. Lett., B, 611, (2005); M. R. Setare, JCAP {\bf 0701}, 023,
(2007); M. R. Setare, Phys. Lett. {\bf B642}, 421, (2006); M. R.
Setare, Phys. Lett. {\bf B644}, 99, (2007).
\bibitem {easther} R. Easther and D. A. Lowe hep-th/9902088.
\bibitem{Hsu} S. D. H. Hsu, Phys. Lett. B, 594, 13, (2004).
\bibitem{varw} U. Alam, V. Sahni, T. D. Saini, A. A. Starobinsky,
Mon. Not. Roy. Astron. Soc., 354, 275 (2004); D. Huterer and A.
Cooray, Phys. Rev., D, 71, 023506, (2005), Y. Wang and M. Tegmark,
astro-ph/0501351.
\bibitem{7}A. Sen, JHEP 0204, 048 (2002); JHEP 0207, 065
(2002); Mod. Phys. Lett. A 17, 1797 (2002); arXiv: hep- th/0312153.
\bibitem{8}A. Sen, JHEP 9910, 008 (1999); E. A. Bergshoeff, M. de Roo,
T. C. de Wit, E. Eyras, S. Panda, JHEP 0005, 009 (2000); J. Kluson,
Phys. Rev. D 62, 126003 (2000); D. Kutasov and V. Niarchos, Nucl.
Phys. B 666, 56, (2003).
\bibitem{fro}A. Frolov, L. Kofman and A. Starobinsky, Phys. Lett.
{\bf B545}, 8, (2002).
\bibitem{gori}V. Gorini, A. Kamenshchik, U. Moschella
and V. Pasquier, and A. Starobinsky, astro-ph/0504576.
\bibitem{setc}M. R. Setare, Phys. Lett. {\bf B648}, 329, (2007).
 \bibitem{nonflat}Q.~G.~Huang and M.~Li,
 JCAP {\bf 0408}, 013 (2004).
 \bibitem{set1}M. R. Setare, Phys. Lett. {\bf B642}, 1, (2006).
\bibitem{set2}M. R. Setare, Jingfei Zhang, Xin Zhang,  JCAP, {\bf 0703}, 007, (2007).
\bibitem{odi1}S. Nojiri,  and S. D. Odintsov, Gen. Rel. Grav. {\bf 38}, 1285,
(2006).
\bibitem{odi2}S. Capozziello, S. Nojiri,  and S. D. Odintsov,  Phys. Lett. {\bf B632}, 597, (2006);
S. Nojiri,  and S. D. Odintsov, hep-th/0611071; S. Nojiri, S. D.
Odintsov, and H. Stefancic, Phys. Rev. {\bf D74}, 086009, (2006).
\bibitem{mset}M. R. Setare, Eur. Phys. J. {\bf C50}, 991, (2007).
\bibitem{cg}
E. J. Copeland, M. R. Garousi, M. Sami, and S. Tsujikawa, Phys. Rev.
{\bf D71}, 043003, (2005).
\bibitem{zhang}J. Zhang, X. Zhang,
H. Liu,  arXiv:0706.1185 [astro-ph], accepted for publication in
Phys. Lett. {\bf B}, (2007).
\bibitem{HG}  Q. G. Huang, Y. Gong, JCAP, 0408, (2004),006.
\bibitem{tac}T. Padmanabhan, Phys. Rev. D66, 021301 (2002); J. S. Bagla, H. K. Jassal and
T. Padmanabhan, Phys. Rev. D67, 063504 (2003); A. Feinstein, Phys.
Rev. D66, 063511, (2002) ;L. R.W. Abramo and F. Finelli, Phys. Lett.
B 575 (2003) 165; J. M. Aguirregabiria and R. Lazkoz, Phys. Rev. D
69, 123502 (2004); Z. K. Guo and Y. Z. Zhang, JCAP 0408, 010 (2004).
\bibitem{cmb1} H. C. Kao, W. L. Lee and F. L. Lin,
 astro-ph/0501487.
\bibitem{cmb3}  J. Shen, B. Wang, E. Abdalla and R. K. Su,
 hep-th/0412227.
\end{thebibliography}
\end{document}